# Technical Report:
# Optimistic Execution in Key-Value Store

Duong Nguyen
*Michigan State University*
nguye476@cse.msu.edu

Aleksey Charapko
*University at Buffalo, SUNY*
acharapk@buffalo.edu

Sandeep Kulkarni
*Michigan State University*
sandeep@cse.msu.edu

Murat Demirbas
*University at Buffalo, SUNY*
demirbas@buffalo.edu

*Abstract*—Limitations of the CAP theorem imply that if availability is desired in the presence of network partitions, one must sacrifice sequential consistency, a consistency model that is more natural for system design. We focus on the problem of what a designer should do if he/she has an algorithm that works correctly with sequential consistency but is faced with an underlying key-value store that provides a weaker (e.g., eventual or causal) consistency. We propose a detect-rollback based approach: The designer identifies a correctness predicate, say $P$, and continues to run the protocol, as our system monitors $P$. If $P$ is violated (because the underlying key-value store provides a weaker consistency), the system rolls back and resumes the computation at a state where $P$ holds.

We evaluate this approach with practical graph applications running on the Voldemort key-value store. Our experiments with deployment on Amazon AWS EC2 instances shows that using eventual consistency with monitoring can provide a $50-80\%$ increase in throughput when compared with sequential consistency. We also show that the overhead of the monitoring itself is low (typically less than 4%) and the latency of detecting violations is small. In particular, more than $99.9\%$ of violations are detected in less than 50 milliseconds in regional AWS networks, and in less than 5 seconds in global AWS networks.

## I. INTRODUCTION

Distributed key-value data stores have gained an increasing popularity due to their simple data model and high performance [1]. A distributed key-value data store, according to CAP theorem [2], [3], cannot simultaneously achieve sequential consistency and availability while tolerating network partitions. Since fault-tolerance, especially the provision of an acceptable level of service in the presence of node or channel failures, is a critical dependability requirement of any system, network partition tolerance is a necessity. Hence, it is inevitable to make trade-offs between availability and consistency, resulting in a spectrum of weaker consistency models such as causal consistency and eventual consistency [1], [4]–[9].

Weaker consistency models are attractive because they have the potential to provide higher throughput and higher customer satisfaction. On the other hand, weaker consistency models suffer from data conflicts. Although such data conflicts are infrequent [1], such incidences will affect the correctness of the computation and invalidate subsequent results.

Furthermore, developing algorithms for the sequential consistency model is easier than developing those for weaker consistency models. Moreover, since sequential consistency model is *more natural*, the designer may already have access to an algorithm that is correct only under sequential consistency. Thus, in this case, the question for the designer is what to do *if the underlying system provides a weaker consistency* or *if the underlying system provides better performance under weaker consistency models*?

As an illustration of such a scenario, consider a distributed computation that relies on a key-value store to arrange exclusive access to a critical resource for the clients. If the key-value store employs sequential consistency and the clients use Peterson's algorithm, mutual exclusion is guaranteed [10], but the performance would be hurt due to the communication overhead of sequential consistency. If eventual consistency is adopted, then mutual exclusion is violated.

In this case, the designer has two options: (1) Either develop a brand new algorithm that works under eventual consistency, or (2) Run the algorithm by pretending that the underlying system satisfies sequential consistency but monitor it to detect violations of the mutual exclusion requirement. In case of the first option, we potentially need to develop a new algorithm for every consistency model used in practice, whereas in case of the second option, the underlying consistency model is irrelevant although we may need to rollback the system to an earlier state if a violation is found. While the rollback in general distributed systems is a challenging task, existing approaches have provided rollback mechanisms for key-value stores with low overhead [11].

The predicate $P$ to monitor depends on the application. For the mutual exclusion application we alluded to above, $P$ might be exclusive access to the shared resource. As another example, consider the following. For many distributed graph processing applications, the clients process a given set of graph nodes. Since the state of a node depends on its neighbors, the clients need to coordinate to avoid updating two neighboring nodes simultaneously. In this case, predicate $P$ is the conjunction of smaller predicates proscribing the concurrent access to some pairs of neighboring nodes (Note that pairs of neighboring nodes belonging to the same client do not need monitoring). The application will continue executing as long as predicate $P$ is true. If $P$ is violated, it will be rolled back to an earlier correct state from where subsequent

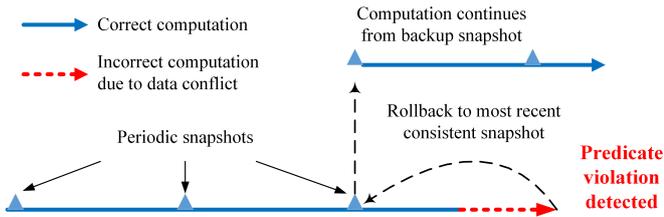

Fig. 1. The detect-rollback approach: when the predicate of interest is violated, system state is restored to the most recent consistent snapshot and the computation resumes from there.

execution will resume (cf. Figure 1).

For performant execution, we require that the monitoring module is non-intrusive, i.e., it allows the underlying system to execute unimpeded. To evaluate the effectiveness of the monitors, we need to identify three parameters: (1) benefit of using the monitors instead of relying on sequential consistency, (2) overhead of the monitors, i.e., how the performance is affected when we introduce the monitors, and (3) detection latency of the monitors, i.e., how long the monitors take to detect violation of $P$. (Note that since the monitoring module is non-intrusive, it cannot prevent violation of $P$.)

**Contributions of the paper.** We implement a monitoring module prototype for the Voldemort key-value data store and run experiments on the Amazon AWS EC2 instances. For the module, we develop monitoring algorithms for linear and semilinear predicates based on the algorithms in [12]–[14]. Our algorithms use Hybrid Vector Clock [15] to help save resources from examining false positive cases thanks to its loosely synchronization with physical clock [16]. We evaluate the monitoring module by running practical graph applications such as social media analysis and weather monitoring on this systems. Besides these experiments, we design some synthetic test cases and setup a local lab network where we can control network condition in order to evaluate the monitoring module in some other aspects such as the impact of workload characteristics and network latency. The observations from the experiments are as follows:

- On Amazon EC2 network, we run the social network analysis application both on sequential consistency without the monitoring module and on eventual consistency with the monitoring module. We observe that –even with the overhead of the monitor– eventual consistency achieves a throughput $50\%$ to $80\%$ higher than that of sequential consistency. Furthermore, in those experiments, we find that violation of mutual exclusion is very rare. On average, one violation every $4,500$ seconds. Hence, the cost of predicate detection and state rollback is outweighed by the benefit of a boosted throughput while the reliability of the computation is still preserved.

- We also evaluate the overhead of the monitoring module if it is intended solely for debugging or runtime monitoring. We find that when the monitors are used with sequential consistency, the overhead is at most $8\%$, even when the monitors are stressed. And, for eventual consistency, the overhead is less than $4\%$.

- Regarding the detection latency of the module, more than $99.9\%$ of violations are detected within 50 milliseconds for Amazon EC2 regional network, and within 5 seconds for global network. In all cases, the latencies are within seventeen seconds.

**Organization of the paper:** Section II we describe the architecture of the key-value store used in this paper. In section III, we define the notion of causality and identify how uncertainty of event ordering in distributed system affects the problem of predicate detection. Section IV describes the overall architecture of the system using monitors. Section V explains the structure of the predicate detection module used in this paper. Section VI presents experimental results and discussions. Section VII compares our paper with related work and we conclude in Section VIII.

## II. SYSTEM ARCHITECTURE

### A. Distributed Key-Value Store

We utilize the standard architecture for key-value stores. Specifically, the data consists of (one or more) tables with two fields, an unique key and the corresponding value. The field value consists of a list of $< version, value >$ pairs. A version is a vector clock that describes the origin of the associated value. It is possible that a key has multiple versions when different clients issue PUT (write) requests for that key independently. When a client issues a GET (read) request for a key, all existing versions of that key will be returned. The client could resolve multiple versions for the same key on its own or use the resolver function provided from the library. To provide efficient access to this table, it is divided into multiple partitions. Furthermore, to provide redundancy and ease of access, the table is replicated across multiple replicas.

To access the entries in this table, the client utilizes two operations, GET and PUT. The operation GET($x$) provides the client the value associated with key $x$. And, the operation, PUT($x, val$), changes the value associated with key $x$ to $val$. The state of the servers can be changed only by PUT requests from clients.

### B. Voldemort Key Store

Voldemort is LinkedIn's open source equivalence of Amazon's Dynamo key-value store. In Voldemort, clients are responsible for handling replication. When connecting to a server for the first time, a client receives meta-data from the server. The meta-data contains the list of servers and their addresses, replication factor ($N$), required reads ($R$), required writes ($W$), and other configuration information.

When a client wants to perform a PUT (or GET), it sends PUT (GET) requests to $N$ servers and waits for the responses for a predefined amount of time (timeout). If at least $W$ ($R$) acknowledgements (responses) are received before the timeout, the PUT (GET) operation is considered successful. If not, the client performs one more round of requests to other servers to get the necessary number of acknowledgements (responses). After the second round, if still less than $W$ ($R$) replies are received, the PUT (GET) operation is considered unsuccessful.

Since the clients do the task of replication, the values $N$, $R$, $W$ specified in the meta-data is only a suggestion. The clients can tune those values for their needs. By adjusting the value of $W$, $R$, and $N$, client can tune the consistency model. For example, if $W + R > N$ and $W > \frac{N}{2}$ for every client, then they obtain sequential consistency. On the other hand, if $W + R \leq N$ then they have eventual consistency.

## III. THE PROBLEM OF PREDICATE DETECTION IN DISTRIBUTED SYSTEMS

Each process execution in a distributed system results in changing its local state, sending messages to other processes or receiving messages from other processes. In turn, this creates a partial order among local states of the processes in distributed systems. This partial order, happened-before relation [17], is defined as follows:

Given two local states $a$ and $b$, we say that $a$ happened before $b$ (denoted as $a \rightarrow b$) iff

- $a$ and $b$ are local states of the same process and $a$ occurred before $b$,
- There exists a message $m$ such that $a$ occurred before sending message $m$ and $b$ occurred after receiving message $m$, or
- There exists a state $c$ such that $a \rightarrow c$ and $c \rightarrow b$.

We say that states $a$ and $b$ are concurrent (denoted as $a \| b$) iff $\neg(a \rightarrow b) \ \wedge \ \neg(b \rightarrow a)$

The goal of a predicate detection algorithm is to ensure that the predicate of interest $P$ is always satisfied during the execution of the distributed system. In other words, we want monitors to notify us of cases where predicate $P$ is violated.

To detect whether the given predicate $P$ is violated, we utilize the notion of *possibility* modality [18], [19]. In particular, the goal is to find a set of local states $e_1, e_2, ..e_n$ such that

- One local state is chosen from every process,
- All chosen states are pairwise concurrent.
- The predicate $\neg P$ is true in the global state $\langle e_1, e_2, \cdots, e_n \rangle$

### A. Vector Clocks and Hybrid Vector Clocks

To determine whether state $a$ happened before state $b$, we can utilize vector clocks or hybrid vector clocks. Vector clocks, defined by Fidge and Mattern [20], [21], are designed for asynchronous distributed systems that make no assumption about underlying speed of processes or about message delivery. Hybrid vector clocks [15] are designed for systems where clocks of processes are synchronized within a given synchronization error (parameter $\epsilon$). While the size of vector clock is always $n$, the number of processes in the system, hybrid vector clocks have the potential to reduce the size to less than $n$.

Our predicate detection module can work with either of these clocks. For simplicity, we recall hybrid vector clocks (HVC) below.

Every process maintains its own HVC. HVC at process $i$, denoted as $HVC_i$, is a vector with $n$ elements such that $HVC_i[j]$ is the most recent information process $i$ knows about the physical clock of process $j$. $HVC_i[i] = PT_i$, the physical time at process $i$. Other elements $HVC_i[j], j \neq i$ is learned through communication. When process $i$ sends a message, it updates its HVC as follows: $HVC_i[i] = PT_i$, $HVC_i[j] = max(HVC_i[j], PT_i - \epsilon)$ for $j \neq i$. Then $HVC_i$ is piggy-backed with the outgoing message. Upon reception of a message $msg$, process $i$ will use the piggy-backed hybrid vector clock $HVC_{msg}$ to update its HVC: $HVC_i[i] = PT_i$, $HVC_i[j] = max(HVC_{msg}[j], PT_i - \epsilon)$ for $j \neq i$.

Hybrid vector clocks are vectors and can be compared as usual. Given two hybrid vector clock $HVC_i$ and $HVC_j$, we says $HVC_i$ is smaller than $HVC_j$, denoted as $HVC_i < HVC_j$, iff $HVC_i[k] \leq HVC_j[k] \forall k$ and $\exists l : HVC_i[l] < HVC_j[l]$. If $\neg(HVC_i < HVC_j) \wedge \neg(HVC_j < HVC_i)$, then the two hybrid vector clocks are concurrent, denoted as $HVC_i \| HVC_j$.

If we set $\epsilon = \infty$, then hybrid vector clocks have the same properties as vector clocks. If $\epsilon$ is finite, certain entries in $HVC_i$ can have the default value $PT_i - \epsilon$ that can be removed. For example, if $n = 10, \epsilon = 20$, a hybrid vector clock $HVC_0 = [100, 80, 80, 95, 80, 80, 100, 80, 80, 80]$ could be represented by $n(10)$ bits 10010010001 and a list of three integers $100, 95, 100$, instead of a list of ten integers.

We use HVC in our implementation to facilitate its use when the number of processes is very large. However, in the experimental results we ignore this optimization and treat as if $\epsilon$ is $\infty$.

### B. Different Types of Predicate Involved in Predicate Detection

In the most general form, predicate $P$ is an arbitrary boolean function on the global state and the problem of detecting $\neg P$ is NP-complete [14]. However, for some classes of predicates such as linear predicates, semilinear predicates, and bounded sum predicates, there exist efficient detection algorithms [12]–[14]. In this paper, we adapt these algorithms for monitoring in key-value stores. Since the correctness of our algorithms follows from the existing algorithms, we omit detailed discussion of the algorithms and focus on their effectiveness in key-value stores.

## IV. A FRAMEWORK FOR OPTIMISTIC EXECUTION

The overall framework for optimistic execution in key-value store is as shown in Figure 2. In addition to the actual system execution in the key-value store, we include local detectors for every server (cf. Figure 4). These local detectors provide information to the monitors. Note that the desired predicate $P$ can be a conjunction of several smaller predicates and each monitor is designed to ensure that a smaller predicate, says $P_i$, continues to be true during the execution. In other words, it is checking if a consistent snapshot where $\neg P_i$ is true (thus $\neg P$ is true) exists.

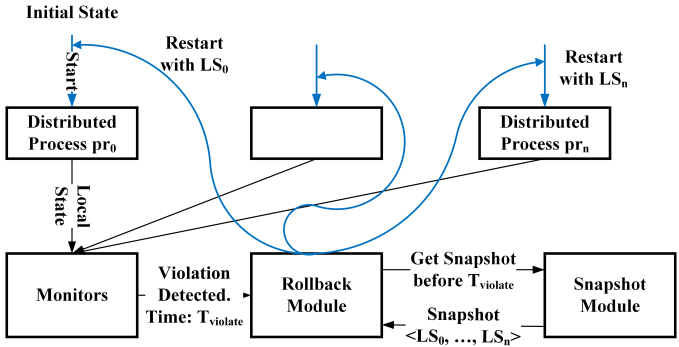

Fig. 2. An overall framework for optimistic execution in key-value store

When the monitors detect violation of the desired property $P$, they notify the rollback module. The monitors also identify a safe estimate of the start time $T_{violate}$ at which the violation occurred, based on the timestamps of local states they received.

If violation of predicate $P$ is rare and the overall system execution is short, we could simply restart the computation from the beginning.

If the system computation is long, we can take periodic snapshots. Hence, when a violation is found, the rollback module notifies all clients and servers to stop the subsequent computation until the restoration to a checkpoint before $T_{violate}$ is complete. The exact length of intervals between the periodic snapshots would depend upon the cost of taking the snapshot and the probability of violating predicate $P$ in the intervals between snapshots.

In case the violations are frequent, feedbacks from the monitor can help the clients to adjust accordingly. For example, if Voldemort clients are running in eventual consistency and find that their computations are restored too frequently, they can switch to sequential consistency by tuning the value of $R$ and $W$ without the involvement of the servers (Recall that in the Voldemort key-value store, the clients are responsible for replication).

Alternatively, we can utilize approach such as Retroscope [11]. Retroscope allows us to dynamically create a consistent snapshot that was valid just before $T_{violate}$ if $T_{violate}$ is within its window-log. This is possible if the predicate detection module is effective enough to detect the violation promptly. In [11], it authors have shown that it is possible to enable rollback for up to 10 minutes while keeping the size of logs manageable.

The approach in Retroscope can be further optimized by identifying the cause of the rollback. For example, consider the example from the Introduction that considers a graph application and requires that two clients do not operate on neighboring nodes simultaneously. Suppose a violation is detected due to clients $C_1$ and $C_2$ operating on neighboring nodes $V_1$ and $V_2$. In this case, we need to rollback $C_1$ and $C_2$ to states before they operated on $V_1$ and $V_2$. However, clients that do not depend upon the inconsistent values of nodes $V_1$ and $V_2$ need not be rolled back.

The goal of this paper is to evaluate the effectiveness of the monitor. In particular, our goal is to determine the overhead of such a monitor and the benefit one could get by running the algorithm with a weaker consistency model. Since this benefit is independent of the strategy used for rollback, we only focus on the effectiveness and overhead of the monitor. With this motivation, the properties of interest in this paper are

- How much overhead occurs when monitors are introduced? This will help us analyze the overhead when monitors are intended for debugging.
- How does the performance of the system compare under the sequential consistency model (where $P$ was guaranteed to be true as the algorithm is correct) with the performance under a weaker consistency model with the monitors?
- How frequent are violations of $P$? This would identify the strategy that is suitable for roll back.
- How long does it take to detect violation of $P$? This would help determine whether logs would be sufficient to provide rollback using approaches such as those in [11].

## V. MONITORING MODULE

The monitoring module is responsible for monitoring and detecting violation of the global predicate of interest in a distributed system. The structure of the module is as shown in Figure 4. It consists of local predicate detectors attached to each server and the monitors independent of the servers. The local predicate detector caches the state of its host server and sends information to the monitors. This is achieved by intercepting the PUT requests for variables that may affect the predicate being monitored. The monitors run predicate detection algorithm based on the information received to determine if the global predicate of interest $P$ has been violated.

We anticipate that the predicate of interest $P$ is a conjunction of all constraints that should be satisfied during the execution. In other words, $P$ is of the form $P_1 \wedge P_2 \wedge \cdots P_l$ where each $P_i$ is an independent predicate and can be of different types (such as linear or semilinear). The job of the monitoring module is to identify an instance where $P$ is violated, i.e., to determine if there is a consistent cut where $\neg P_1 \vee \neg P_2 \vee \cdots \neg P_l$ is true. In order to monitor multiple predicates, the designer can have multiple monitors with one monitor for each predicate $P_i$ or one monitor for all predicates $P_i$'s. In the former case, the detection latency is small but the overheads can be unaffordable when the number of predicates is large since we need many monitor processes. In the latter case, the overhead is small but the detection latency is long. We adopt a compromise: our monitoring module consists of multiple monitors and each monitor is responsible for multiple predicates. The predicates are assigned to the monitors based on the hash of the predicate names in order to balance the monitors' workload.

The number of monitors equals the number of servers and the monitors are distributed among the machines running the

```
<predicate>
  <type>semilinear</type>
  <conjClause>
    <id>0</id>
    <var>
      <name>x2</name> <value>1</value>
    </var>
    <var>
      <name>y2</name> <value>1</value>
    </var>
  </conjClause>
  <conjClause>
    <id>1</id>
    <var>
      <name>z2</name> <value>1</value>
    </var>
  </conjClause>
</predicate>
```

Fig. 3. XML specification for $\neg P \equiv (x_1 = 1 \wedge y_1 = 1) \vee z_2 = 1$

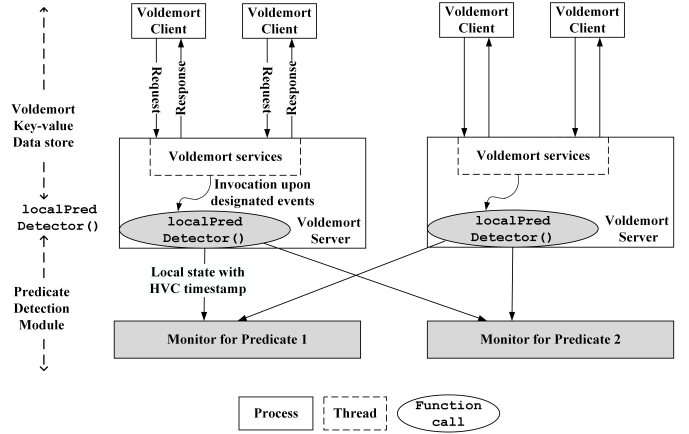

Fig. 4. Architecture of predicate detection module

servers. We have done so to ensure that the cost of the monitors is accounted for in experimental results while avoiding overloading a single machine. An alternative approach is to have monitors on a different machine. In this case, the trade-off is between CPU cycles used by the monitors (when monitors are co-located with servers) and communication cost (when monitors are on a different machine). Our experiments suggest that in the latter approach (monitors on a different machine) monitoring is more efficient. However, since there is no effective way to compute the increased cost (of machines in terms of money), we report results where monitors are on the same machines as the servers.

Note that in Figure 4, each monitor is depicted as one process. In implementation, a monitor may consists of multiple distributed processes collaborating to monitor a single (smaller) predicate. For simplicity, each monitor is one process in this paper discussion.

Each (smaller) predicate $P_i$ is a boolean formula on the states of some variables. Since any boolean formula can be converted to disjunctive normal form, users can provide the predicates being detected ($\neg P_i$'s) in disjunctive normal form. We use XML format to represent the predicate. For example, the semilinear predicate, says $\neg P_1 \equiv (x_1 = 1 \wedge y_1 = 1) \vee z_2 = 1$, in XML format is shown in Figure 3. Observe that this XML format also identifies the type of the predicate (linear, semilinear, etc.) so that the monitor can decide the algorithm to be used for detection.

**Implementation of Local Predicate Detectors.** Upon the execution of a PUT request, the server calls the interface function `localPredicateDetector` which examines the state change and sends a message (also known as a candidate) to one or more monitors if appropriate. Note that not all state changes cause the `localPredicateDetector` to send candidates to the monitors. The most common example for this is when the changed variable is not relevant to the predicates being detected. Other examples depend upon the type of predicate being detected. As an illustration, if predicate $\neg P$ is of the form $x_1 \wedge x_2$ then we only need to worry about the case where $x_i$ changes from $false$ to $true$.

The local predicate detector maintains a cache of variables related to the predicates of interest to efficiently monitor the server state. A candidate sent to the monitor of predicate $P_i$ consists of an HVC interval and a partial copy of server local state containing variables relevant to $P_i$. The HVC interval is the time interval on the server when $P_i$ is violated, and the local state has the values of variables which make $\neg P_i$ true.

For example, assume the global predicate of interest to be detected is $\neg P \equiv \neg P_1 \vee \neg P_2 \cdots \vee \neg P_m$ where each $\neg P_j$ is a smaller global predicate. Assume that monitor $M_j$ is responsible for detection of predicate $\neg P_j$. Consider any smaller predicate, says $\neg P_2$, and for the sake of the example, assume that it is a conjunctive predicate, i.e. $\neg P_2 \equiv (\neg LP_2^1) \wedge (\neg LP_2^2) \wedge ...(\neg LP_2^n)$ where $n$ is the number of servers. We want to detect when $\neg P_2$ becomes true. On a server, says server $i$, the local predicate detector will monitor the corresponding local predicate $\neg LP_2^i$ (or $\neg LP_2$ for short, in the context of server $i$ as shown in Figure 5). Since $\neg P_2$ is true only when all constituent local predicates are true, server $i$ only has to send candidates for the time interval when $\neg LP_2$ is true. In Figure 5, upon the first PUT request, no candidate is sent to monitor $M_2$ because $\neg LP_2$ is false during interval $[HVC_i^0, HVC_i^1]$. After serving the first PUT request, the new local state makes $\neg LP_2$ true, starting from the time $HVC_i^2$. Therefore upon the second PUT request, a candidate is sent to monitor $M_2$ because $\neg LP_2$ is true during the interval $[HVC_i^2, HVC_i^3]$. Note this candidate transmission is independent of whether $\neg LP_2$ is true or not after the second PUT request is served. It depends on whether $\neg LP_2$ is true after execution of the previous PUT request. That is why, upon the second PUT request, a candidate is also sent to monitor $M_3$ but none is sent to $M_1$. Note that if the predicate is semi-linear, then upon a PUT request for a relevant variable, the local predicate detector has to send a candidate to the associated monitor anyway.

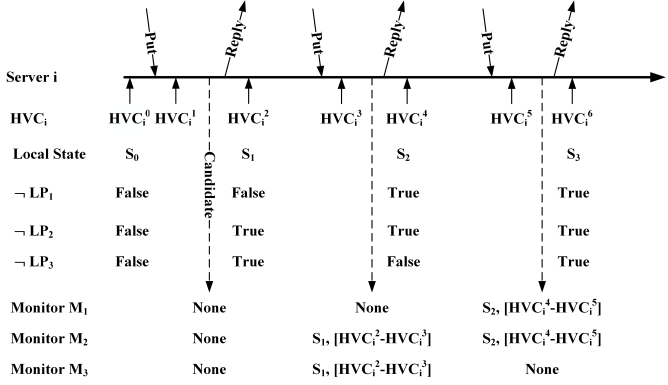

| Server i | | | | | | | |
|---|---|---|---|---|---|---|---|
| HVC$_i$ | HVC$_i^a$ HVC$_i^b$ | HVC$_i^c$ | HVC$_i^d$ | HVC$_i^e$ | HVC$_i^f$ | HVC$_i^g$ | |
| Local State | S$_0$ | | S$_1$ | | S$_2$ | | S$_3$ |
| $\neg$LP$_1$ | False | | False | | True | | True |
| $\neg$LP$_2$ | False | | True | | True | | True |
| $\neg$LP$_3$ | False | | True | | False | | True |
| Monitor M$_1$ | | None | | None | | S$_2$, [HVC$_i^e$-HVC$_i^h$] | |
| Monitor M$_2$ | | None | | S$_1$, [HVC$_i^c$-HVC$_i^h$] | | S$_2$, [HVC$_i^e$-HVC$_i^h$] | |
| Monitor M$_3$ | | None | | S$_1$, [HVC$_i^c$-HVC$_i^h$] | | None | |

Fig. 5. Illustration of candidates sent from a server to monitors corresponding to three conjunctive predicates. If the predicate is semilinear, the candidate is always sent upon a PUT request of relevant variables.

**Implementation of the monitors.** The task of a monitor is to determine if some smaller predicate $P_i$ under its responsibility is violated, i.e., to detect if a consistent state on which $\neg P_i$ is true exists in the system execution. The monitor constructs a global view of the variables relevant to $P_i$ from the candidates it receives. The global view is valid if all candidates in the global view are pairwise concurrent.

The concurrence/causality relationship between a pair of candidates is determined as follows: suppose we have two candidates $Cand_1, Cand_2$ from two servers $S_1, S_2$ and their corresponding HVC intervals $[HVC_1^{start}, HVC_1^{end}], [HVC_2^{start}, HVC_2^{end}]$. Without loss of generality, assume that $\neg(HVC_1^{start} > HVC_2^{start})$ (cf. Figure 6).

- If $HVC_2^{start} < HVC_1^{end}$ then the two intervals have common time segment and $Cand_1 \| Cand_2$.
- If $HVC_1^{end} < HVC_2^{start}$, and $HVC_1^{end}[S_1] \leq HVC_2^{start}[S_2] - \epsilon$ then interval one is considered happens before interval two. Note that $HVC[i]$ is the element corresponding to process $i$ in HVC. In this case $Cand_1 \rightarrow Cand_2$
- If $HVC_1^{end} < HVC_2^{start}$, and $HVC_1^{end}[S_1] > HVC_2^{start}[S_2] - \epsilon$, this is the uncertain case where the intervals may or may not have common segment. In order to avoid missing possible bugs, the candidates are considered concurrent.

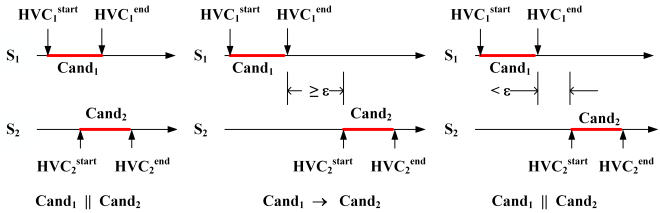

Fig. 6. Illustration of causality relation under HVC interval perspective

When a global predicate is detected, the monitor informs the administrator or triggers a designated process of recovery. We

**Algorithm 1** Monitor algorithm for linear predicate

1: **Input:**
2:    $P$        ▷ global linear predicate to monitor
3: **Variable:**
4:    $GS$                      ▷ global state
5: **Initialization:**
6:    $GS \leftarrow$ set of initial local states
7: **while** P(GS)==true **do**
8:    Find forbidden local state $s \in GS$
9:    $GS \leftarrow GS \cup succ(s)$    ▷ advance $GS$ along $s$
10:    consistent($GS$)    ▷ make $GS$ consistent
11: **end while**
12: **return** $GS$

**Algorithm 2** Monitor algorithm for semilinear predicate

1: **Input:**
2:    $P$      ▷ global semilinear predicate to monitor
3: **Variable:**
4:    $GS$                      ▷ global state
5: **Initialization:**
6:    $GS \leftarrow$ set of initial local states
7: **while** P(GS)==true **do**
8:    Find a local state $s \in GS$ such that $s \in eligible(GS)$ and $s$ a semi-forbidden state of $P$ in $GS$.
9:    $GS \leftarrow GS \cup succ(s)$    ▷ advance $GS$ along $s$
10: **end while**
11: **return** $GS$

develop detection algorithms for the monitors of linear predicates and semilinear predicates based on [13], [14] as shown in Algorithm 1 and Algorithm 2. Basically, the algorithms have to identify the correct candidates to update the global state ($GS$) so that we would not have to consider all possible combinations of $GS$ as well as not miss the possible violations. In linear (or semilinear) predicates, these candidates are forbidden (or semi-forbidden) states. Forbidden states are states such that if we do not replace them, we would not be able to find the violation. Therefore, we must advance the global state along forbidden states. Semi-forbidden states are states such that if we advance the global state along them, we would find a violation if there exists any. When advancing global state along a candidate, that candidate may not be concurrent with other candidates existing in the global state. In that case, we have to advance the candidates to make them consistent. This is done by `consistent(GS)` in the algorithm. If we can advance global state along a candidate without `consistent(GS)`, that candidate is called an eligible state. The set of all eligible states in global state is denoted as `eligible(GS)` in the algorithms. For more detailed discussion of linear and semilinear predicates, we refer to [14].

After a consistent global state $GS$ is obtained, we evaluate whether predicate $P$ is violated at this global state ($P(GS) = true$ means $P$ is satisfied, $P(GS) = false$ means

$P$ is violated). If $P$ is violated, the algorithms return the global snapshot $GS$ as the evidence of violation. Note that the monitors will keep running even after a violation is reported so that possible violations in the future will not be missed. This is the case when the applications, after being informed about the violation and rolling back to a consistent checkpoint before the moment when the violation occurred, continue their execution and violations occur again due to faults (e.g. network delay). Hence the monitors have to keep running in order detect any violations of $P$.

The way we evaluate $P$ on global state $GS$ is slightly different from the algorithms in [12]–[14], [22]. In those algorithms, the candidates are sent directly from the clients containing the states of the clients. In our algorithms, the candidates are sent from the servers containing the information the servers know about the states of the clients that have been committed to the store by the clients. Note that, in a key-value store, the clients use the server store for sharing variables and committing updates. Therefore, the states of clients will eventually be reflected at the server store. Since the predicate $P$ is defined over the states of the clients, in order to detection violations of $P$ from the states stored at the server, we have to adapt the algorithms in [12]–[14], [22] to consider that difference. Furthermore, the state of a client can be stored slightly differently at different servers. For example, an PUT request may be successful at the regional server but not successful at remote servers. In that case, assuming we are using eventual consistency, the regional server store will have the update while remote stores do not have the update. Our algorithms also consider this factor when evaluating $P$.

Since our algorithms are adapted from [12]–[14], [22], the correctness of our algorithms follow from those existing algorithms. For more detailed discussion and proof of correctness of the algorithms, please refer to [12]–[14], [22].

**Handling a large number of predicates.** When the number of predicates to be monitored is large (e.g. hundreds of thousands, as in *Social Media Analysis* application in next section or in graph-based applications discussed in the Introduction), it is costly to maintain monitoring resources (memory, CPU cycles) for all of them simultaneously. That not only slows down the detection latency but also consumes all the resources on the machines hosting the monitors (for example, we received `OutOfMemoryError` error when monitoring tens of thousands predicates simultaneously). However, we observe that not all predicates are active at the same time. Only predicates relevant to the nodes that the clients are currently working on are active. A predicate is considered inactive when there is no activity related to that predicate for a predetermined period of time, and therefore the evaluation of that predicate is unchanged. The monitors will clean up resources allocated for that predicate to save memory and processing time.

**Automatic inference of predicate from variable names.** This feature is also motivated by applications where the number of predicates to be monitored is large such as the graph-based applications. In this case, it is impossible for the users to manually specify all the predicates. However, if the variables relevant to the predicates follow some convention, our monitoring module can automatically generate predicates on-demand. For example, in graph-based applications, the predicates are the mutual exclusion on any edge whose end-points are assigned to two different clients. Let $A\_B$ is such an edge, and assume $A < B$. If the clients are using Peterson's mutual exclusion, the predicate for edge $A\_B$ will be

$$\neg P_{A\_B} \equiv (flagA\_B\_A = true \land turnA\_B = "A")$$
$$\land (flagA\_B\_B = true \land turnA\_B = "B")$$

When a server receives a request (PUT or GET) from some client for a variable whose name is either `flagA_B_A`, or `flagA_B_B`, or `turnA_B`, it knows that the client is interested in the lock for edge $A\_B$ and the server will generate the predicate for edge $A\_B$ so that the monitors can detect if the mutual exclusion access on edge $A\_B$ is violated. On the other hand, if the servers never see requests for variables `flagA_B_A`, `flagA_B_B`, and `turnA_B`, then both nodes $A$ and $B$ are assigned to the same client and we do not need the mutual exclusion predicate for edge $A\_B$.

## VI. Evaluation Results and Discussion

### A. Experimental Setup

**System Configurations.** We run experiments on Amazon AWS EC2 instances. The servers run on M5.xlarge instances (4 vCPUs and 16 GB RAM). The clients run on M5.large instances (2 vCPUs and 8 GB RAM). The EC2 instances locate in three AWS regions: Ohio, U.S; Oregon, U.S; Frankfurt, Germany.

We also run experiments on our local lab network which is set up so that we can control network latency. We use 9 commodity PCs, 3 for servers, 6 for clients, with configurations as in Table I. Each client machine hosts multiple client processes, while each server machine host one Voldemort server process. We control the delay by placing proxies between the clients and the servers. For all clients on the same physical machine, there is one proxy process for those clients. All communication between those clients and any server, as shown in Figure 7, is relayed through that proxy. Due to the proxy delays, machines are virtually arranged into three regions as in Figure 8. Latency within a region is small (2 $ms$) while those between regions are high and tunable (e.g. 50, 100 $ms$). Since Voldemort uses active replication, we do not place proxies between servers.

TABLE I
Machine configuration in local lab experiments

| Machine | CPU | RAM |
|---|---|---|
| Server machine 1, 2 | 4 Intel Core i5 3.33 GHz | 4 GB |
| Server machine 3 | 4 Intel Core i3 3.70 GHz | 8 GB |
| Client machine 1, 2 | 4 Intel Core i5 3.33 GHz | 4 GB |
| Client machine 3, 4 | Intel Core Duo 3.00 GHz | 4 GB |
| Client machine 5 | 4 AMD Athlon II 2.8 GHz | 6 GB |
| Client machine 6 | 4 Intel Core i5 2.30 GHz | 4 GB |

We considered replication factors ($N$) of 3 and 5. The parameters $R$ (required reads) and $W$ (required writes) are

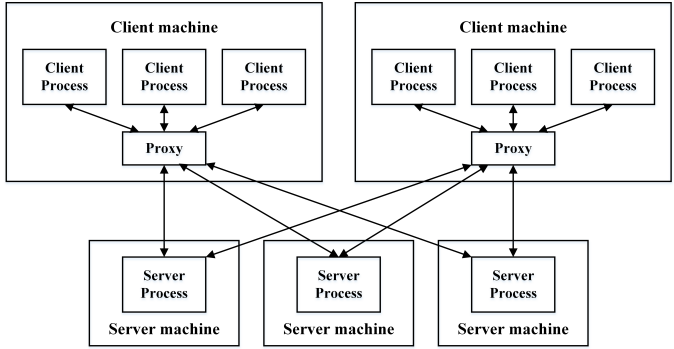

Fig. 7. Simulating network delay using proxies

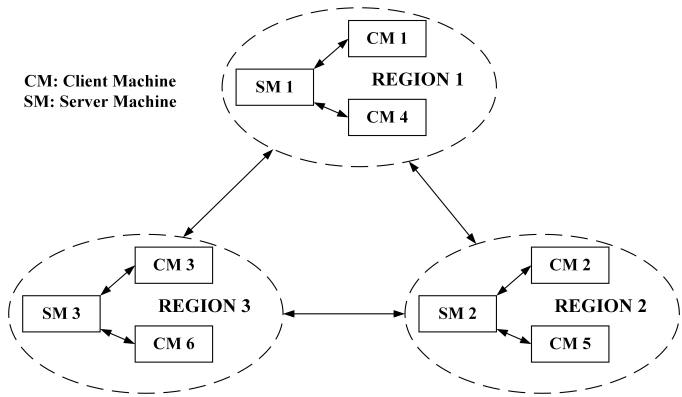

CM: Client Machine
SM: Server Machine

Fig. 8. Network arrangement with proxies

| N | R | W | Abbreviation | Consistency model |
|---|---|---|---|---|
| 3 | 1 | 3 | N3R1W3 | Sequential |
|   | 2 | 2 | N2R2W2 | Sequential |
|   | 1 | 1 | N3R1W1 | Eventual |
| 5 | 1 | 5 | N5R1W5 | Sequential |
|   | 3 | 3 | N5R3W3 | Sequential |
|   | 1 | 1 | N5R1W1 | Eventual |

chosen to achieve different consistency models as shown in Table II. The number of servers is equal to the replication factor $N$. The number of clients is varied between 15 and 90.

**Test cases.** In our experiments, we use 3 case studies: *Social Media Analysis*, *Weather Monitoring*, and *Conjunctive*.

The *Social Media Analysis* application considers a large graph representing users and their connections. The goal of clients is to update the state of each user (node) based on its connections. For the sake of analysis in our analysis, the attribute associated with each user is a color and the task is to assign each node a color that is different from its neighbors. The input graph is generated by the tool *networkx* that simulates the power-law degree distribution and the clustering characteristics of social networks. The graph has 50,000 nodes with about 150,000 edges. Each client is assigned a set of nodes to be colored and run a distributed coloring algorithm [23].

Since the color of a node is chosen based on its neighbors' colors, while a client $C_1$ is coloring node $v_1$, no other client is updating the colors of $v_1$'s neighbors. The goal of the monitors is to detect violation of this requirement. This requirement can be viewed as a mutual exclusion (semi-linear) predicate where a client going to update the color of $v_1$ has to obtain exclusive locks associated with all edges incident to $v_1$. Mutual exclusion is guaranteed if clients use Peterson's algorithm and the system provides sequential consistency [10]. However, it may be violated in eventual consistency model. To avoid deadlock, clients obtain locks in a consistent order. For example, let $A\_B$ be the lock associated with edge between nodes $A$ and $B$, and $A < B$. Then lock $A\_B$ is obtained before $C\_D$ when $A < C$ or when $A = C$ and $B < D$.

Since the graph contains some nodes of very high degree, it is far sub-optimal if we use $maxDeg$ (the maximum degree in graph) colors. Preprocessing high degree nodes reduces the number of colors used by the algorithm as well as the number of locks sought by the clients. A node is considered high degree if its degree is greater than the threshold $q$. The value of $q$ is chosen such that the number of nodes with degree greater than $q$ is smaller than $q$. So high degree nodes could be colored by at most $q$ colors and the whole graph would use no more than $2q$ colors. In our experiment with various size graphs similar to our input graph, the number of nodes with degree $deg$ roughly follows the distribution

$$count(deg) \approx 6.5 \times |V| \times deg^{-2.5}$$

where $|V|$ is the number of nodes. Therefore the number of nodes with degree greater than $q$ is

$$gCount(q) \approx \int_{q+1}^{|V|} count(deg)ddeg$$

By solving $q >= gCount(q)$ we roughly have:

$$q \gtrapprox (\frac{11 \times |V|}{3})^{\frac{1}{2.5}}$$

For example, $|V| = 50000$, without (and with) preprocessing high degree nodes, the algorithm uses 1650 (and 255) colors.

The number of predicates being monitored in this test cases is proportional to the number of edges, even after preprocessing high degree nodes which account for less than 10% of edges.

We note that the task performed by each client (i.e., choosing the color of a node) is just used as an example. It is easily generalizable for other analysis of social media graph (e.g., finding clusters, collaborative learning, etc.)

The *Weather Monitoring* application considers a planar graph where the state of each node is affected by the state of its neighbors. In this application, we model a client that updates the state of each node by reading the state of its neighbors and updating its own state. This application can be tailored to vary the ratio of GET/PUT request. This application can be generalized into practical planar graph problem such as weather forecasting [24], radiocoloring in wireless and sensor network [25], computing Voronoi diagram [26].

Finally, the *Conjunctive* application is an instance of distributed debugging where the predicate being detected (i.e., $\neg P$) is of the form $P_1 \wedge P_2 \wedge \cdots \wedge P_l$. Each local predicate $P_i$ becomes true with a probability $\beta$ and the goal of the monitors is to determine if the global conjunctive predicate $\neg P$ becomes true. In this application, we monitor multiple conjunctive predicates simultaneously. Since we can control how frequently these predicates become true, we can use it mainly to assess monitoring latency and stress test the monitors. Conjunctive predicates are useful in distributed testing such as to specify breakpoints.

**Performance Metric and Measurement.** We use throughput as the performance metrics in our experiments. Throughput can be measured at two perspective: application, and Voldemort server. The two perspectives are not the same but related. One application request triggers multiple requests at Voldemort client. For example one application PUT request is translated into one GET_VERSION request (to obtain the last version of the key) and one PUT request (with a new incremented version) at the Voldemort client library. Then each Voldemort client request causes multiple requests at servers due to replication. Failures and timeout also make the counts differ. For example, an application request is served and counted at server but if the server response is lost or arrives after timeout, the request is considered unsuccessful and thus not counted at the application. Generally, servers' counts are greater than applications' counts. In our experiment, we use the aggregated measurement at servers to assess the overhead of our approach since the monitors directly interfere with the servers operation, and use aggregated measurement at applications to assess the benefit of our approach because that measurement is close to users' perspective. Hence, in the following sections, for the same experiment, please be aware that the measurements used for overhead and benefit evaluation are different.

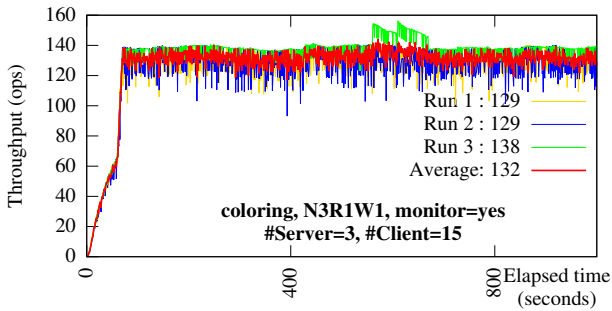

Fig. 9. Illustration of result stabilization. The application *Social Media Analysis* (coloring) is run three times on AWS with monitoring enabled. Number of servers ($N$) = 3. Number of clients per server ($C/N$) = 5. Aggregated throughput measured by *Social Media Analysis* application in three different runs and their average is shown.

**Results stabilization.** For each experiment, we run three times and use the average as the representative results for that experiment. Figure 9 shows the stabilization of different runs in several configurations. Note that the values are aggregated from all applications (or servers). We observe that in every run, after a short period of initialization, the measurements converge on a stable value. When evaluate our approach, we use the values measured at the stable phase. We also note that the aggregated throughput in 9 is not very high but expected. The pairwise round-trip latency between three AWS regions (Ohio, Oregon, Frankfurt) are 76, 103, and 163 $ms$. The average round-trip latency is 114 $ms$. We will roughly estimate the cost of a GET request since in *Social Media Analysis*, most operations are GET requests to read lock availability and colors of neighbors. A GET request is executed by Voldemort client in two steps:

1. Perform parallel request: client simultaneously send GET requests to all servers (N = 3) and wait for responses with timeout of 500 $ms$. The wait is over when either client gets responses from all servers or the timeout expires. In this case, client will get all responses in about 117 $ms$ (114 $ms$ for communication delay, and 3 $ms$ for server processing time).

2. Perform serial request: client checks if it has received enough required responses. If not, it has to send addition GET requests to servers to get enough number of responses. If after the additional requests, the required number of responses is not met, the GET request is considered unsuccessful. Otherwise, result is returned. In the current case, the number of responses received (3) is greater than the required (R = 1). Thus this step is skipped.

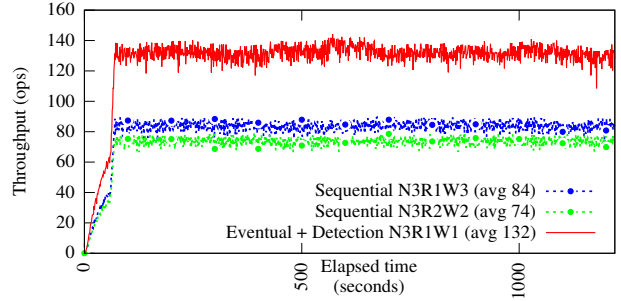

Fig. 10. Benefit of eventual consistency with monitors vs. sequential consistency without monitors in the *Social Media Analysis* application on AWS environment. Number of servers = 3. Number of clients = 15. The throughput improvement compared to R1W3 and R2W2 is 57% and 78%, respectively.

Hence a GET request takes roughly 117 $ms$ to complete, on average. Since GET is the dominating operation in the *Social Media Analysis* application, with 15 clients, the expected aggregated throughput is $\frac{15}{0.117} = 128\ ops$. Note if the latency decreases the aggregated throughput will increase (cf. Figure 12).

### B. Experimental Results on Amazon AWS

**Comparison of sequential consistency and monitors with eventual consistency.** As discussed in the introduction, one of the problems faced by the designers is that they have access to an algorithm that is correct under sequential consistency but the underlying key-value store provides a weaker consistency. In this case, one of the choices is to pretend as if sequential consistency is available but monitor the critical predicate $P$. If this predicate is violated, we need to rollback to an earlier state and resume the computation from there. Clearly, this approach would be feasible if the monitored computation with eventual consistency provides sufficient benefit compared with sequential consistency. In this section, we evaluate this benefit.

Figures 10 compares the performance of our algorithms for eventual consistency with monitors and sequential consistency without monitors in the *Social Media Analysis* application on the AWS environment. Using our approach, the performance is boosted to 57% (for N3R1W3) and 78% (for N3R2W2). Note that the cost of a GET request is more expensive in N3R2W2 (the required number of positive acknowledgement is 2) than in N3R1W3 (the required acknowledgement is 1). Since in the *Social Media Analysis* application GET requests dominates, the application performs better in N3R1W3 than in N3R2W2.

**Overhead of monitoring.** A weaker consistency allows the application boost performance on key-value store as illustrated above. To ensure correctness, a weaker consistency needs monitors to detect violations and trigger rollback recovery when such violations happen. As a separate tool, the monitors are useful in debugging to ensure that the program satisfies the desired property throughout the execution. In all cases, it is desirable that the overhead of the monitors is small so that they would not curtail the benefit of weaker consistency or enlarge the debugging cost significantly.

Figure 11 shows the overhead of the monitoring on different consistency models in the *Social Media Analysis* application. The overhead is between 1% and 2%. At its peek, the number of active predicates being monitored reaches 20,000 predicates. Thus, the overhead remains reasonable even with monitoring many predicates simultaneously.

**Violation detection and recovery.** In our experiment with *Social Media Analysis* applications on eventual consistency (N3R1W1), in many executions of total 9,000 seconds, we detect only 2 instances of mutual exclusion violations. Those violations are detected within 2,238 $ms$ and 2,213 $ms$ since the time the violations occur. So for *Social Media Analysis* application, violations could happen on eventual consistency every 4,500 $s$ on average. This rate is small enough so that the cost of general rollback recovery is outweighed by the performance benefit of eventual consistency.

**Discussion.** Since network is generally good, eventual consistency is reliable while providing significantly higher performance than sequential consistency does. Data conflicts could happen but very rare, as shown in our experiment and in [1]. Therefore the cost of recovery from inconsistent state is small compared to the benefit of our approach. Furthermore, in some graph applications such as *Social Media Analysis*, recovery could be achieved efficiently as follows. We have the clients process graph nodes in tasks (or batches). Each task (batch) contains some configurable number of nodes, for example 10 nodes. Since nodes have different connectivity, the processing time for each task varies. In our experiments, the minimum, average, and maximum processing time for tasks of size 10 are 22,645 $ms$, 45,136 $ms$, and 217,369 $ms$, respectively. Since the time of processing a task is far beyond the time to detect violation, if violation occurs (and is detected), the client just needs to abort and restart the current task. Furthermore, *Social Media Analysis* clients can defer updating the colors of nodes in current task until the end of the task before committing the updates. If no violation is reported, clients issue PUT requests to update nodes' colors normally. If a violation is reported, they just skip those PUT requests and restart the task again or change the task. Note that this abort and restart mechanism does not involve any state rollback at the servers.

For *Social Media Analysis* application, the procedure of aborting and restarting can be efficiently achieved follows. *Social Media Analysis* clients can choose to defer updating the colors of nodes in current task until the end of the task. If no violation is reported, it will issue PUT requests to update nodes' colors normally. If a violation is reported, it just aborts those PUT requests and restarts the task again. Note that this abort and restart mechanism for *Social Media Analysis* clients does not involve any state rollback at the servers.

**Impact of workload characteristics.** In order to evaluate the impact of workload on our algorithms we run the *Weather Monitoring* application where the proportional of PUT and GET is configurable. The number of servers is 5 and the number of clients is 10. The machines hosting the servers and clients are in the same AWS region (North Virginia, U.S.) but in 5 different availability zones (North Virginia is the only region with more than 4 different availability zones). We choose machines in the same region to reduce the latency (to less than 2 $ms$), thus increasing the throughput measure and stressing the servers and the monitors. If we choose machines in regions similar to the experiments with *Social Media Analysis* above, in order to achieve the same level of aggregated throughput of this section (cf. Figure 12, we would have to increase the number of clients 100 times.

From Figures 12(a) and 12(b), we find that when the percentage of PUT request increases from 25% to 50%, the benefit over sequential consistency (N5R1W5 in this case) increases from 18% to 37%. This is because the cost for a PUT request is expensive in N5R1W5 as a PUT request is successful only when it is confirmed by all 5 servers. Thus, when the proportion of PUT increases, the performance of N5R1W5 decreases. In such cases, sequential settings that balance R and W (e.g. N5R3W3) will perform better than settings emphasize W (e.g. N5R1W5). When GET requests dominate, it is vice versa (cf. Figure 10). We also observe that, when PUT percentage increases and other parameters are unchanged, the aggregated throughput measured at clients decreases. That is because a PUT request consists of a GET_VERSION request (which is as expensive as a GET request) and an actual PUT request, therefore PUT request takes longer time to complete than a GET request does.

Regarding overhead, Figure 12(c) shows that the overhead is 4% when PUT percentage is 50%. Note that in *Weather Monitoring* application, the number of predicates being monitored is proportional to the number of clients. Thus, the overhead remains reasonable even with monitoring several predicates simultaneously and servers are stressed.

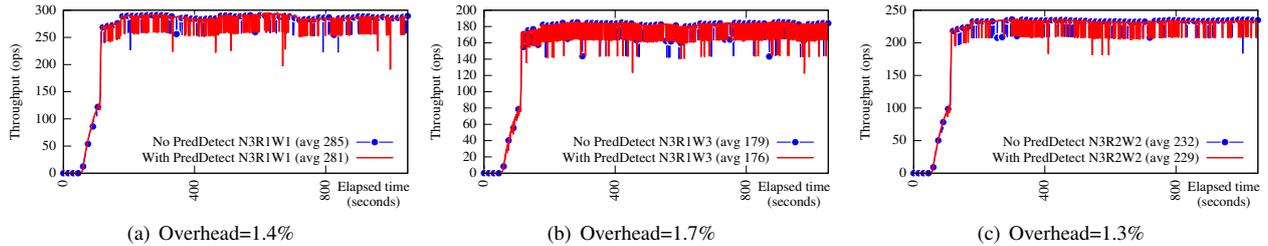

Fig. 11. . The overhead of of running monitors on different consistency levels, in *Social Media Analysis* application. Number of servers = 3. Number of clients = 15.

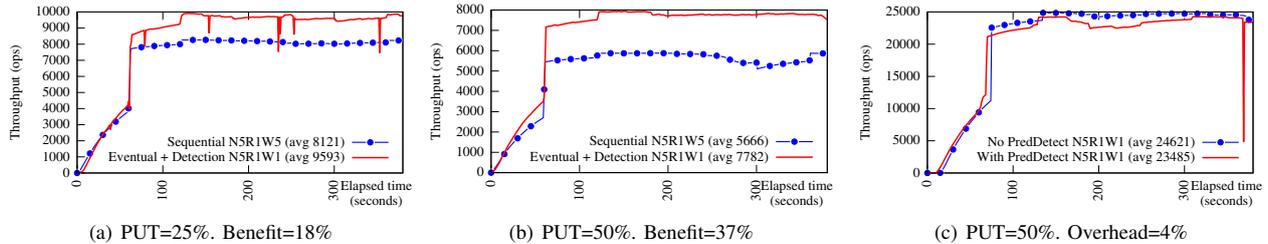

Fig. 12. Benefit and overhead of monitors in *Weather Monitoring* application. Percentage of PUT requests is 25% and 50% Number of servers =5. Number of clients = 10. Machines are on the AWS North Virginia region but in different availability zones.

The number of violations detected in this experiment is only one instance in executions with total time of $18,000$ $ms$. The violation is detected within $20$ $ms$. This observation again supports that although violation of mutual exclusion in eventual consistency is a theoretical possibility, it is quite rare. In order to evaluate the detection latency of monitors with higher statistical reliability, we need test cases where violations are more frequent.

TABLE III
RESPONSE TIME IN 20,647 CONJUNCTIVE PREDICATE VIOLATIONS

| Response time (milliseconds) | Count | Percentage |
|---|---|---|
| < 50 | 20,632 | 99.927% |
| 50 − 1,000 | 6 | 0.029% |
| 1,000 − 10,000 | 3 | 0.015% |
| 10,000 − 17,000 | 6 | 0.029% |

**Detection latency** is the time elapsed between violation of the predicate being monitored and the time when the monitors detect it. In this experiment, clients will run *Conjunctive* application in the same AWS configuration as *Weather Monitoring* application above. The monitors have to detect violations of conjunctive predicates of the form $P = P_1 \wedge P_2 \wedge \cdots P_{10}$. Furthermore, we can control how often these predicates become true by randomly changing when local predicates are true. In these experiments, the rate of local predicate being true ($\beta$) is $1\%$, which is chosen based on the time breakdown of some MapReduce applications [27], [28]. The PUT percentage is 50%. The *Conjunctive* application is designed so that the number of predicate violations is large and to stress the monitors. We considered both eventual consistency and sequential consistency. Table III shows detection latency distribution of more than $20,000$ violations recorded in the *Conjunctive* experiments. Predicate violations are generally detected promptly. Specifically, $99.93\%$ of violations were detected in $50$ $ms$, $99.97\%$ of violations were detected in $1$ $s$. There are rare cases where detection latency is greater than ten seconds. Among all the runs, the maximum detection latency recorded is 17 seconds, the average is $8$ $ms$.

Regarding overhead and benefit, the overhead of monitors on N5R1W1, N5R1W5, and N5R3W3 is 7.81%, 6.50%, and 4.66%, respectively. The benefit of N5R1W1 over N5R1W5 and N5R3W3 is 27.90% and 20.16%, respectively. Observing Figures 12(c), we find that there are a few moments where the aggregated throughput of all servers drops down. This is happening because some or all servers are spending a significant computation for the local predicate detection module. Such moments are infrequent. Furthermore, since each M5.large server used in our experiment has only two Voldemort server threads, when one of the thread is running the predicate detection module, the aggregated throughput would be clearly affected. In a typical setting, each server runs a large number of server threads. Thus, when a server process is running predicate detection module, the decrease in aggregated throughput would be less noticeable.

In results presented so far, we have considered experiments under various factors but network latency. In next section, we consider impact of network latency on the monitors.

| Latency (ms) | Application | Client/ Server | Monitor | N3R1W1 | | | N3R2W2 | | | | N3R1W3 | | | |
|---|---|---|---|---|---|---|---|---|---|---|---|---|---|---|
| | | | | server | overhead | app | server | overhead | app | benefit | server | overhead | app | benefit |
| 50 | Conjunctive | 20 | yes | 821 | -0.2% | 470 | 842 | 0.6% | 375 | 25.3% | 588 | 3.3% | 337 | 40.7% |
| | | | no | 819 | | 470 | 847 | | 375 | | 608 | | 334 | |
| | Weather Monitoring | 20 | yes | 924 | 0.2% | 454 | 795 | 7.1% | 345 | 27.2% | 628 | 3.2% | 312 | 45.0% |
| | | | no | 926 | | 453 | 856 | | 357 | | 649 | | 313 | |
| | Social Media Analysis | 10 | yes | 560 | 0.2% | 258 | 367 | 0.5% | 156 | 65.4% | 344 | 7.8% | 174 | 47.4% |
| | | | no | 561 | | 267 | 369 | | 156 | | 373 | | 175 | |
| 100 | Conjunctive | 20 | yes | 476 | 0.4% | 270 | 491 | -0.2% | 218 | 23.3% | 354 | 0.0% | 191 | 42.1% |
| | | | no | 478 | | 271 | 490 | | 219 | | 354 | | 190 | |
| | Weather Monitoring | 20 | yes | 544 | 0.7% | 266 | 500 | 1.0% | 209 | 28.5% | 371 | 0.8% | 176 | 49.4% |
| | | | no | 548 | | 273 | 505 | | 207 | | 374 | | 178 | |
| | Social Media Analysis | 10 | yes | 287 | 0.0% | 135 | 236 | 0.0% | 74 | 80% | 185 | -0.5% | 86 | 60.7% |
| | | | no | 287 | | 133 | 236 | | 75 | | 184 | | 84 | |

## C. Experimental Results on Local Lab Network

**Latency distribution.** The latency between clients and servers is controlled by proxies in local lab settings. The latency is simulated to follow Gamma distribution [29], [30]. With Gamma distribution, the latency between two nodes $A$ and $B$ is $D_{A,B} = D_{A,B}^d + D_{A,B}^s$. $D_{A,B}^d$ is the deterministic delay between $A$ and $B$ reflecting the topological distance between the two nodes. If the path between two nodes is fixed then $D_{A,B}^d$ is assumed to be a constant. $D_{A,B}^d$ is in the range of $[50, 100]$ [1]. $D_{A,B}^s$ is the stochastic delay reflecting the network condition between $A$ and $B$ at the time of transmission. $D_{A,B}^s$ is assumed to observe a Gamma distribution with shape factor between 0.6 and 1.0 [29]. In our experiments, we calculate $D_{A,B}^s = sample \times multiplier$ where $sample$ is a sample from Gamma distribution with shape factor of 0.8, and $multiplier$ is the multiplier to convert scalar value to latency unit (i.e. milliseconds). We choose $multiplier = D_{A,B}^d \times 0.2$ by experiments. So

$$D_{A,B} = D_{A,B}^d \times (1 + sample \times 0.2)$$

**Impact of latency.** In this experiment, the one-way latency within a region (cf. Figure 8) is 1 ms and one-way latency between regions varies from 50 ms to 100 ms. In Table IV, the overhead is computed by comparing server measurements when the monitors are enabled and disabled, the benefit is computed by comparing application measurements on sequential consistency without monitoring to those on the corresponding eventual consistency with monitoring. For example, when one-way latency is 50 *ms*, if we run the *Weather Monitoring* application on N3R1W3, the overhead of monitoring is $(649 - 628)/649 = 3.2\%$. If we run the same application on eventual consistency N3R1W1 with monitoring, the benefit (compared to running on N3R1W3 without monitoring) is $(454 - 313)/313 = 45\%$.

From Table IV, as latency increases, the benefit of eventual consistency and monitoring vs. sequential consistency increases. For example when one-way latency increases from 50 ms to 100 ms, in *Social Media Analysis* application, the benefit of eventual consistency and monitoring vs. sequential

---

[1] According to https://www.cloudping.co/, the round trip latency between AWS regions could be between 40 ms to 300 ms.

---

consistency R1W3 increases from 47% to 60%. In case of R2W2, the increase is from 65% to 80%. This increase is expected because when latency increases, the chance for a request to be successful at a remote server decreases. Due to strict replication requirement of sequential consistency, client will have to repeat the request again. On the other hand, on eventual consistency, requests are likely to be successfully served a local server and the client can continue regardless of results at remote servers. Hence, as servers are distributed in more geographically disperse locations, the benefit of eventual consistency is more noticeable. Regarding overheads, they are generally less than 4%. In all cases, the overheads are at most 8%.

## VII. RELATED WORK

### A. Predicate Detection in Distributed Systems

Predicate detection is an important task in distributed debugging. An algorithm for capturing consistent global snapshots and detecting stable predicates was proposed by Chandy and Lamport [31]. A framework for general predicate detection is introduced by Marzullo and Neiger [19] for asynchronous systems, and Stollers [18] for partially synchronous systems. These general frameworks face the challenge of state explosion as the predicate detection problem is NP-hard in general [14]. However, there exist efficient detection algorithms for several classes of practical predicates such as unstable predicates [22], [32], [33], conjunctive predicates [13], [34], linear predicates, semilinear predicates, bounded sum predicates [14]. Some techniques such as partial-order method [35] and computation slicing [36], [37] are also approaches to address the NP-Completeness of predicate detection. Those works use vector clocks to determine causality and the monitors receive states directly from the constituent processes. Furthermore, the processes are static. [38], [39] address the predicate detection in dynamic distributed systems. However, the classes of predicate is limited to conjunctive predicate. In this paper, our algorithms are adapted for detecting the predicate from only the states of the servers in key-value store, not from the clients. The servers are static (except failure), but the clients can be dynamics. The predicates supported include linear (including conjunctive) predicates and semilinear predicates.

We use hybrid vector clocks to determine causality in our algorithms. In [16], the authors discussed the impact of various factors, among which is clock synchronization error, on precision of predicate detection module. In this paper, we set epsilon at a safe upper bound for practical clock synchronization error to avoid missing potential violations. In other words, hybrid vector clock is practically vector clock. Furthermore, this paper focuses on the efficiency and effectiveness of predicate detection module.

### B. Distributed data-stores

Many NoSQL data-stores exist on the market today, and a vast portion of these systems provide eventual consistency. The eventual consistency model is especially popular among key-value and column-family databases. The original Dynamo [1] was one of pioneers in the eventual consistency movement and served as the basis for Voldemort key-value store. Dynamo introduced the idea of hash-ring for data-sharding and distribution, but unlike Voldemort it relied on server-side replication instead of active client replication. Certain modern databases, such as Cosmos DB and DynamoDB [40], [41] offer tunable consistency guarantees, allowing operators to balance consistency and performance. This flexibility would enable some applications to take advantage of optimistic execution, while allowing other applications to operate under stronger guarantees if needed. However, many data-stores [42], [43] are designed to provide strong consistency and may not benefit from optimistic execution module.

Aside from general purpose databases, a variety of specialized solutions exist. For instance, TAO [44] handles social graph data at Facebook. TAO is not strongly consistent, as its main goal is performance and high scalability, even across datacenters and geographical regions. Gorilla [45] is another Facebook's specialized store. It operates on performance time-series data and highly tuned for Facebook's global architecture. Gorilla also favors availability over consistency in regards to the CAP theorem.

### C. Snapshots and Reset

The problem of acquiring past snapshots of a system state and rolling back to these snapshots has been studied extensively. Freeze-frame file system [46] uses Hybrid Logical Clock (HLC) to implement a multi-version Apache HDFS. Retroscope [11] takes advantage of HLC to find consistent cuts in the system's global state be examining the state-history logs independently on each node of the system. The snapshots produced by Retroscope can later be used for node reset by simple swapping of data-files. Eidetic systems [47] take a different approach and do not record all prior state changes. Instead, eidetic system records any non-deterministic changes at the operating system level and constructing a model to navigate deterministic state mutations. This allows the system to revert the state of an entire machine, including the operating system, data and applications, to some prior point. Certain applications may not require past snapshots and instead need to quickly identify consistent snapshots in the presence of concurrent requests affecting the data. VLS [48] is one such example designed to provide snapshots for data-analytics applications while supporting high throughput of requests executing against the system.

### D. Distributed Data Processing

MapReduce [49], DataFlow [50] are general-purpose distributed data processing frameworks. In the realm of distributed graph processing, many frameworks are available such as Pregel [51], GraphLab [52], GraphX [53], and PowerGraph [54]. In those works, data is persisted in semi-structural storages such Google File System, Hadoop Distributed File Systems [55], BigTable [56], or in-memory storage such as Spark [57]. Our work focuses on the no-structure key-value stores and the impact of different consistency models on key-value store performance. Our approach's usefulness is also not limited to graph applications.

## VIII. CONCLUSION

Due to limitations of the CAP theorem and the desire to provide availability/good performance during network partitions (or long network delays), many key-value stores choose to provide a weaker consistency such as eventual or causal consistency. This means that the designers need to develop new algorithms that work correctly under such weaker consistency models. An alternative approach is to run the algorithm by ignoring that the underlying system is not sequentially consistent but monitoring it for violations that may affect the application. For example, in case of graph-based applications (such as those encountered in weather monitoring, social media analysis, etc.), each client operates on a subset of nodes in the graph. It is required that two clients do not operate on neighboring nodes simultaneously. In this case, the predicate of interest is that local mutual exclusion is always satisfied.

We demonstrated the usage of this approach in the Voldemort in cases where we have two types of predicates: conjunctive predicates and semilinear predicates (such as that required for local mutual exclusion). In our experiments on computing distributed graph coloring in Amazon AWS, our approach improve the throughput performance of the computation from 50% to 80%. Furthermore, we find that the number of violation of predicates of interest was very rare. All violations are also detected promptly. In regional network, violations are detected within $50\ ms$ while in global network, they can be detected within $5\ s$. Thus, the amount of work wasted due to rollback would be very small especially if one utilizes techniques such as Retroscope [11] that allows one to roll back the system to an earlier state on-demand. For graph processing applications such as *Social Media Analysis*, if we defer clients' updates until the end of a task, the recovery can be achieved by restarting the task without state rollback.

Furthermore, it is also feasible to utilize these monitors for the sake of debugging as well. In particular, the overhead of the monitors is very low. The overhead is typically less than $4\%$ and in stressed experiments less than $8\%$.

With feedback from the monitors, clients could tune their operation mode accordingly. For example, when the network condition is unstable for an extended period of time, violations reported by monitors are more frequent. In that case, instead of recovering state repeatedly, clients can switch to the sequential consistency model. Note that systems like Voldemort use active replication, therefore the clients can choose the consistency level as they want.

There are several possible future work in this area. This paper considered linear and semilinear predicates. In general, the problem of predicate detection is NP-complete. Hence, we intend to evaluate the practical cost of general predicate detection algorithms. We are also working on making these algorithms more efficient by permitting them to occasionally detect phantom violations. We are evaluating whether this increased efficiency would be worthwhile even though some unnecessary rollbacks may occur. Another future work is to integrate the monitor with Retroscope [11] to automate the rollback and recovery.